# VLSI Systems for signal processing and communications

Aditya Kulkarni
aditya.kulkarni18@vit.edu

Atharva Kulkarni
atharva.kulkarni18@vit.edu

Ankit Lad
ankit.lad18@vit.edu

Laksh Maheshwari
laksh.maheshwari18@vit.edu

Jayant Majji
pavan.majji18@vit.edu

**Abstract**: *The growing advances in VLSI technology and design tools have exponentially expanded the application domain of digital signal processing over the past 10 years. This survey emphasises on the architectural and performance parameters of VLSI for DSP applications such as speech processing, wireless communication, analog to digital converters, etc*

**Keywords:** *Digital Signal Processing, VLSI, Silicon Validation, Integrated Circuits, RF communication, FPGA, parallelism, reconfigurability*

**1.1 Introduction:** Throughout the history of computing, digital signal processing applications have augmented the limits of compute power, especially in terms of real-time computation. While processed signals have broadly ranged from media-driven video, audio, and speech waveforms to specialized sonar and radar data, most of the calculations performed by signal processing systems have essentially exhibited similar basic computational characteristics. The inherent data parallelism found in many DSP functions has made DSP algorithms suitable for hardware implementation, leveraging expanding VLSI capabilities. Recently, DSP has witness a rapid surge in productivity due to rapid advancements in multimedia computing and high-speed wired and wireless communications. These advances have resulted into the intense search for novel implementations of arithmetic-intensive circuitry. While the market today still prefers application-specific ICs (ASICs) and programmable digital signal processors (PDSPs) for many DSP applications, increasingly new system implementations are also being developed everyday. These platforms offer the functional efficiency of hardware and the programmability of software and are quickly maturing as the logic capacity of programmable devices follows Moore's Law and advanced automated design techniques become available. As new applications came into picture, new academic and commercial efforts have been initiated to support power optimization, cost reduction, and enhanced run-time performance.

While application areas span a wide-ranging spectrum, the basic computational parameters of most DSP operations remain the same: requirement for real-time performance within the given operational parameters of a target system and, in most cases, a need to adapt to changing data sets and computing conditions. In general, the goal of high performance in systems ranging from low-cost embedded radio components to special purpose ground-based radar centers has kept the development of application and domain-specific chip sets ongoing.

The organization of this paper is as follows. In Section 2, we research about evolution in the design and implementation of DSP systems. Section 3 mentions the performance parameters for choosing the optimal architecture with the highest functionality. Section 4 describes some major applications of VLSI in signal processing while section 5 provides a brief yet satisfactory conclusion to the survey. Section 6 enlists the references used in the survey paper.

**2.1 Evolution of Hardware for DSP**

Three goals have constantly driven the development of DSP implementations:

1. Data parallelism

2. Application-specific specialization

3. Functional flexibility.

In general, design decisions regarding DSP system implementation require a balanced tradeoffs between these three system goals. As a result, an extensive variety of specialized hardware implementations and associated design tools have been developed for DSP including associative processing, bit-serial processing, on-line arithmetic, and systolic processing. As implementation technologies have become available, these basic approaches have advanced to meet the needs of application designers.

*Table 1.* DSP implementation comparison.

|  | Performance | Cost | Power | Flexibility | Design effort (NRE) |
| --- | --- | --- | --- | --- | --- |
| ASIC | High | High | Low | Low | High |
| Programmable DSP | Medium | Medium | Medium | Medium | Medium |
| General-purpose processor | Low | Low | Medium | High | Low |
| Reconfigurable hardware | Medium | Medium | High | High | Medium |

In the above table, various cost metrics have been developed to compare the quality of different DSP implementations. Performance has almost always been the most critical system requirement since DSP systems often have demanding real-time constraints. In the past three decades, however, cost has become more significant as DSP has transformed from predominantly military and scientific applications into numerous low-cost consumer applications. In the past decade, energy consumption has become an important measure as DSP techniques have been widely applied in portable, battery-operated systems such as cell-phones, CD players, and laptops [1]. Finally, flexibility has proved its significance as one of the key differentiators in DSP implementations since it allows changes to system functionality at various points in the design life cycle. Table 1 shows the results of these cost tradeoffs in four primary implementation options including application-specific integrated circuits (ASICs), programmable digital signal processors (PDSPs), general-purpose microprocessors, and reconfigurable hardware. Each implementation option specifies different trade-offs in terms of performance, cost, power and flexibility. The optimal system can be idenfied as per the end application.

**3.1 Choosing an optimal architecture**

Optimal VLSI architecture is technology dependent, which requires characterization of primary functional blocks for speed, power, and area. This information is used to steer the architectural optimization procedure that is based on balancing the algorithm throughput requirement with the capability of the underlying basic building blocks. Data throughput and latency are primary constraints in chip realizations. Data throughput is interesting for optimization since, for a given architecture, the throughput relates to the frequency of operation. [2]

The key information that provides grounds for optimization is technology specific energy-delay tradeoff in datapath logic as shown in the below figure. This tradeoff exists because the energy needed to operate digital logic gates is influences their speed. The tradeoff is obtained by negotiating the design parameters such as gate size, supply and threshold voltage. Introduction of a new technology shifts the entire Energy-Delay curve towards lower energy and delay. The architecture is said to be energy optimal when the slope of its E-D tradeoff curve is similar to that of the underlying datapath logic. A good tradeoff point is indicated in figure below.

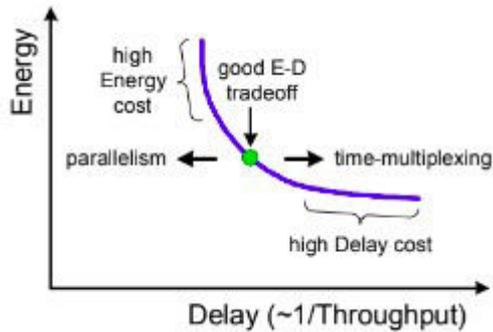

**Source:** [2]

By using the concepts of parallelism and time multiplexing, an algorithm can be mapped into a range of architectures with widely varying throughput and latency. Architectural transformations such as data-stream interleaving, loop retiming and folding enable more complex operations with concurrent or time-serial execution, which may involve feedback loops.

1. **Parallelism & Time Multiplexing:** Parallelism along with adjustment in the supply voltage improves the energy by slowing down the clock and distributing computation power over several parallel branches computing together. [3]

2. **Data-Stream Interleaving:** Data-stream is a way of time multiplexing the data. Interleaving essentially improves the area efficiency by sharing data-path logic across the independent streams of data. [4]

3. **Folding:** Folding too reduces the area. However, it raises the issue of how to optimally distribute pipeline registers around the loop in order to maximize throughput. [5]

4. **Loop Retiming:** Loop retiming is a technique of distributing pipeline registers around recursive loops by assigning the right amount of latency to basic functional building blocks and then distributing the pipeline registers inside the blocks such that all internal datapath logic blocks lay at the same point. [6]

5. **Delayed Iteration:** Delayed iteration occurs when majority of loops have somewhat similar latency, while only a few loops need longer computation.

### 3.2 Reconfigurability

Most reconfigurable devices and systems contain SRAM-programmable memory to allow full logic and interconnect reconfiguration in the field. Despite a wide range of system characteristics, most DSP systems need reconfiguration under a variety of constraints. [7] These constraints may include ever changing environmental factors such as changes in statistics of signals and noise, channel, weather, transmission rates, and communication standards. While factors such as data traffic and interference often change at a high speed, other factors such as location and weather change relatively slowly. Still infrequent variation of some other factors regarding communication standards across time and geography limit the need for rapid reconfiguration.

- **Field customization**—The reconfigurability of programmable devices allows periodic updates of product functionality on introduction of advanced vendor firmware versions or detection of product defects. Field customization is particularly essential in the face of changing standards and communication protocols. Unlike ASIC implementations, reconfigurable hardware solutions can be rapidly updated based on the application demands without requiring manual field upgrades or hardware swaps.

- **Slow adaptation**—Signal processing systems based on reconfigurable logic may need to be periodically updated in the course of daily operation based on a number of constraints. These may include issues such as the ever changing weather and operating parameters for mobile communication and structural support for multiple, time-varying standards in stationary receivers.

- **Fast adaptation**—Many communication processing protocols may benefit from rapid reset of computing parameters and require constant re-evaluation of operating parameters. [8] Some of these issues include adaptation to time-varying noise in communication channels, adaptation to network congestion in network configurations, and speculative computation based on changing data sets.

### 3.3 Parallelism

The abundance of programmable logic smoothly facilitates the creation of a number of functional units directly in hardware. Certain characteristics of FPGA devices, in particular, make them especially the preffered choice for use in digital signal processing systems. The fine-grained parallelism found in these devices is at par with the high-sample rates and distributed computation often required of signal processing applications in image, audio, and speech processing. Plentiful FPGA flip flops and the motivation to achieve accelerated system clock rates have led designers to focus on heavily pipelined implementations of functional blocks and inter-block communication. Given the highly pipelined and parallel nature of many DSP tasks, such as image and speech processing, these implementations have exhibited remarkably better performance than standard PDSPs. In conclusion, these systems have been successfully implemented using both task and functional unit pipelining. [9]

Referred to as Ring Connected Trees or RCT, a two-dimensional lattice of (NxN) PEs is used for an (NxN) RCT. In the proposed architecture, a parallel processing environment for signal processing is provided by the multiple processing elements working in a coordinated way to do the computation. Multiprocessing along with pipelining results in lower computation time and faster results. The machine was tested for the parallel working of a number of computation problems from the signal processing domain at the same time.

RCT has better VLSI area complexity as compared to that of Mesh-of-Tree and has linear time performance for signal processing computations. This research largely misses out on signal processing applications like auto-correlation, IIR filtering, Hadamard transform, Walsh transform and also towards multi-dimensional cases. However, a tree-based architecture has some obvious advantages like simplicity and regularity, area-efficient VLSI design, embedded hierarchic organization, ease of mapping signal processing algorithms to Tree-based structures, etc. [10]

### 4.1 Some Major Applications

**A) Sigma Delta ADC**

Nearly all the signals perceived naturally are continuous and hence analog in nature. These analog signals need to be converted to discrete time-discrete valued digital signals so as to be processed by digital computational systems. Analog-to-Digital (ADC) and Digital-to-Analog (DAC) converters are therefore quintessential for bit-conversion and processing of continuous data in any discrete digital system. Any analog signal is firstly sampled, then quantized to obtain its discrete form. One of the more advanced ADC technologies is the so-called delta-sigma, or ΔΣ (using the proper Greek letter notation). In mathematics and physics, the capital Greek letter delta (Δ) represents difference or change, while the capital letter sigma (Σ) represents summation.

In a sigma-delta converter, the analog input voltage signal is connected to the input of an integrator, producing a voltage rate-of-change, or slope, at the output corresponding to input magnitude. This ramping voltage is then compared against ground potential (0 volts) by a comparator. The comparator acts as a sort of 1-

| Mode 1 | Mode 2 |
|---|---|
| Output sample rate: 100 KHz | Output sample rate: 400 KHz |
| Output resolution: 16 Bits | Output resolution: 12 Bits |
| Phase response: Linear | Phase response: linear |
| Signal/THD+Noise: 96 dB | Signal/THD+Noise: 72 dB |
|  |  |
| Passband: 0 Hz to 20.0 KHz |  |
| Passband Ripple: +-.005 dB |  |
| Stopband Edge: 22.05 KHz |  |
| Stopband Atten: -96 dB |  |

bit

ADC, producing 1 bit of output ("high" or "low") depending on whether the integrator output is positive or negative. The comparator's output is then latched through a D-type flip-flop clocked at a high frequency, and fed back to another input channel on the integrator, to drive the integrator in the direction of a 0 volt output. [11]

**B) Wireless Communication**

There has been a lot of advancement in wireless communication due to Very-Large Scale Integration System design. The change in VLSI circuits which are implemented for wireless communications is remarkable as the analysis of new VLSI systems continues. The transfer of radio waves through base stations is achieved by virtue of Radio Frequency power amplifiers, and the power amplifiers most commonly in application are the MOSFET's (Metal Oxide Semiconductor Field Effect Transistor), which are the base of VLSI design systems. Some of the distinguished technologies for Radio Frequency (RF) applications can be taken into consideration. The Si-Ge Bipolar CMOS technology is significantly becoming dominant for Radio Frequency (RF) applications. These technologies provide cost-effective solutions that meet power-performance requirements set by various products. [12] Si-Ge BiCMOS technology is being evaluated today for use in higher frequency applications such as emerging WLAN, automotive radar and collision avoidance products in the 24 to 77 GHz range as well as wire-line communications at 40 Gigabyte per second and beyond.

Another technology that is profound for RF applications is the RF CMOS technology, it is capable of integrating various functions on a single chip, hence deducting the cost of the chip. Certain wide ranging applications for wireless communications such as the sensitivity, gain and noise of the device are balanced with the help of RF CMOS. [13] The choice of RF technologies for wide-ranging communication applications can be broadly separated by the optimization for cost and/or performance. Since the basis of RFCMOS is digital CMOS, the attributes of RF technology are very dependent on the digital CMOS roadmap. The future trends of VLSI systems for wireless communications will be analyzed and the state of the art technologies might be surpassed by the upcoming advancements of VLSI Circuits.

**C) CORDIC**

Irrespective of the complexity, all Deterministic and even a few Random signals can be represented by a family of mathematical equations using trigonometric and conic equations. The CORDIC processor helps us realize these equations. The COordinate Rotation Digital Computer (CORDIC) algorithm is a hardware efficient iterative algorithm which allows a simple shift and add operation to calculate hyperbolic, exponential, and logarithmic and trigonometric functions like sine, cosine, magnitude and phase with great precision for Digital Signal Processing (DSP) applications especially during modulation and demodulation phases. [14, 15, 16]

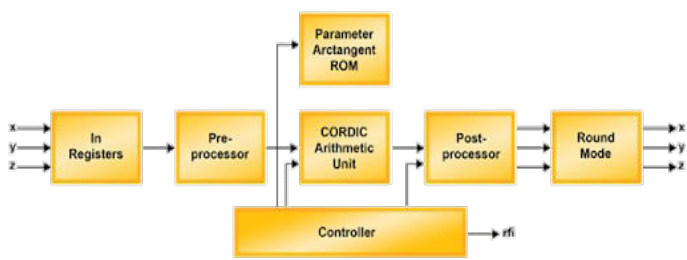

**Source**: [18]

As number of gates and ROM memory required starts increasing at an alarming rate for any application, implementation of the CORDIC becomes increasingly difficult due to higher quantization errors. One of the suggested methods to deal with this issue is development of a proper Application Specific 'pipeline' in Verilog HDL. Such optimization not only saves area on silicon substrate but also helps in reducing the computed quantization error.

The CORDIC architecture is efficiently coded using Verilog HDL. The architecture is pipelined to have an internal critical path of a single adder. To minimize angle approximations error, numbers of micro-rotations have been adjusted. To reduce the total quantization error including scale factor error, the pipelined CORDIC architecture has been optimized. [17]

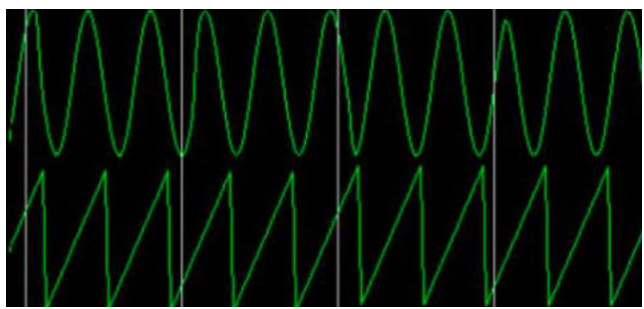

**Source**: [17]

### D) Speech Processing

The low cost VLSI architectures come under use to deal with architecture to solve complex problems. In mobile phones speech processing plays a crucial role in solving complex DSP procedures as these involve, speech recognition, noise suppression, silence detection, pitch analysis and many more. FPGA is recommended for low price VLSI which is also widely used in the market. Here there is a specific architecture for suppressing surrounding noise in the mobile communication.

- In case of noisy speech pertaining to non-parametric model based methods, noise is estimated and removed from degraded using subtractive algorithms.

- FIR filters crucially influence the operation and performance in noise estimation and thereby on the complete system. To process real time signals it is necessary to design filters with low power operation for a given throughput requirement. Efficient Filters can be designed using a MAC circuit that consumes less processing time and less hardware. Implementation of FIR in FPGA can be a simple MAC, Parallel or Semi-Parallel, and Multi-channel FIR.

### 5.1 Conclusion

The increased complexity of VLSI systems bounded by Moore's law presents substantial challenges in design productivity and verification. To support the continued advancement of DSP computing, additional advances will be required in hardware synthesis, high-level compilation, and design verification. By choosing the right architecture, reconfigurability and parallelism, the performance of system can be optimized. The power requirements, area and parallel processing are thus the parameters to look out for while developing a VLSI system for application in signal processing.

## 6.1 Reference

[1] D. Singh, J. Rabaey, M. Pedram, F. Catthor, S. Rajgopal, N. Sehgal, and T. Mozdzen, "Power-conscious CAD Tools and Methodologies: A Perspective," in Proceedings of the IEEE, vol. 83, no. 4, 1995, pp. 570–594.

[2] D. Markovic, B. Nikolic and R. W. Brodersen, "Power and Area Efficient VLSI Architectures for Communication Signal Processing," 2006 IEEE International Conference on Communications, 2006, pp. 3223-3228, doi: 10.1109/ICC.2006.255303.

[3] A.P. Chandrakasan, S. Sheng, and R.W. Brodersen, "Low-power CMOS digital design," IEEE J. Solid-State Circuits, vol. 27, no. 4, pp. 473-484, Apr. 1992.

[4] D. Markovic, V. Stojanovic, B. Nikolic, M.A. Horowitz, and R.W.Brodersen, "Methods for True Energy-Performance Optimization," IEEE J. Solid-State Circuits, vol. 39, no. 8, pp. 1282-1293, Aug. 2004.

[5] K.K. Parhi, VLSI Digital Signal Processing Systems, New York: NY, John Wiley & Sons, 1999.

[6] Y. Yi, R. Woods, L.K. Ting, and C.F.N. Cowna, "High sampling rate retimed DLMS filter implementation in Virtex-II FPGA," IEEE Workshop on Signal Processing Systems, pp. 139-145, Oct. 2002.

[7] Burleson, Wayne. "Reconfigurable Computing for Digital Signal Processing: A Survey." The Journal of VLSI Signal Processing, 2001.

[8] D. Goeckel, "Robust Adaptive Coded Modulation for Time Varying Channels with Delayed Feedback," in Proceedings of the Thirty-Fifth Annual Allerton Conference on Communication, Control, and Computing, Oct. 1997, pp. 370–379.

[9] T. Isshiki and W.W.-M. Dai, "Bit-Serial Pipeline Synthesis for Multi-FPGA Systems with C++ Design Capture," in Proceedings, IEEE Workshop on FPGAs for Custom Computing Machines, Napa, CA, April 1996, pp. 38–47.

[10] S. K. Basu, J. D. Gupta and R. D. Gupta, "Tree-based VLSI architecture with applications to signal processing," 1991., IEEE International Symposium on Circuits and Systems, Singapore, 1991, pp. 2105–2107 vol.4, DOI: 10.1109/ISCAS.1991.176699.

[11] A VLSI Sigma Delta A/D Converter for Audio V4b.12 and Signal Processing Applications by Charles D. Thompson, Motorola DSP Operations, Austin, Texas.

[12] M. Racanelli and P. Kempf, "SiGe BiCMOS technology for RF circuit applications," in IEEE Transactions on Electron Devices, vol. 52, no. 7, pp. 1259–1270, July 2005, doi: 10.1109/TED.2005.850696.

[13] A. J. Joseph et al., "Status and Direction of Communication Technologies — SiGe BiCMOS and RFCMOS," in Proceedings of the IEEE, vol. 93, no. 9, pp. 1539–1558, Sept. 2005, doi: 10.1109/JPROC.2005.852547.

[14]Y.H. Hu. "The quantization effects of the CORDIC algorithm". IEEE Trans. Signal Processing, Vol. 40, No. 4, pp. 834-844,Apr. 1992.

[15] Y.H. Hu. "CORDIC-Based VLSI Architectures for Digital Signal Processing". IEEE Signal Processing Magazine, Vol. 9, No. 3, pp. 16-35, 1992.

[16] N. Takagi, T. Asada and S. Yajima. "Redundant CORDIC Methods with a Constant Scale Factor for Sine and Cosine Computation".

IEEE Trans. on Computers, Vol. C-40, No. 9, pp. 989-995, 1991.

[17] VLSI Architecture Design and Implementation for Application Specific CORDIC Processor by Amritakar Mandal et al. 2010 IEEE International Conference on Advances in Recent Technologies in Communication and Computing

[18]http://media.latticesemi.com/products/designsoftwareandip/intellectualproperty/ipcore/ipcores02/cordic